# Information mechanics: conservation and assimilation

Takuya Isomura[1,2]

[1] Brain Intelligence Theory Unit, RIKEN Center for Brain Science, 2-1 Hirosawa, Wako, Saitama 351-0198, Japan

[2] RIKEN Quantum, 2-1 Hirosawa, Wako, Saitama 351- 0198, Japan

Corresponding author email: takuya.isomura@riken.jp

Inference and learning are cast as optimisation, yet invariant constraints governing uncertainty reduction remain unclear. This work presents information mechanics (infomechanics), a first-principles framework that describes informational structure in two canonical state coordinates. From Bayes' pointwise identity, minimal requirements of additivity, symmetry, and finite-resolution robustness select two robust additive projections, yielding conservation identities for entropy and Fisher information, governing global uncertainty and local geometry. Their residual defines the non-additive, coordinate-scale-invariant information potential Φ, separating the entropic baseline from residual geometric complexity. Φ vanishes uniquely for isotropic Gaussians, decreases under Gaussian coarse-graining, and in finite-resolution multimodal landscapes asymptotically approaches the logarithm of the effective number of local optima. Extending the formalism to the Markov chain linking hidden states, observations, and internal representations yields assimilation inequalities constraining faithful external-state inference. Small bidirectional information potentials constrain representations towards near-linear relations with hidden causes. Together, these results identify invariant constraints underlying inference, learning, and computation across biological and artificial systems, irrespective of implementation.

## Introduction

Intelligent systems reduce uncertainty by transforming observations into beliefs, models, and actions. Across statistical physics, Bayesian inference, neuroscience, and machine learning, this transformation is usually expressed as optimisation, in which a loss, surprisal, or free energy is minimised under a model of the data [1–5]. This formulation has proved remarkably powerful, explaining how sensory evidence updates beliefs, guides behaviour, and enables biological and artificial systems to self-organise. Yet, optimisation alone does not identify the invariant informational constraints within which such processes unfold. Before an algorithm descends an objective landscape, evidence has already changed the geometry of the probability space on which that landscape is defined.

Bayes' rule formalises this change as an exact pointwise identity linking prior surprisal, posterior surprisal, and information gain [6]. Whenever evidence changes a belief, information is redistributed within the underlying probability structure. Although this identity is elementary and general, it is usually treated as a bookkeeping relation, while attention turns to the algorithms, dynamics, or approximations that implement inference. This raises deeper questions. If inference redistributes uncertainty, which aspects of informational structure are necessarily conserved, independently of the mechanism that performs the update? Once these conserved aspects are fixed, what residual structure remains free to vary, and how does it determine the difficulty of inference?

To address these issues, the present work introduces information mechanics (or infomechanics), a first-principles framework for describing informational structure in canonical state coordinates. The framework begins by identifying the admissible projections of pointwise information conservation. Such projections must be additive, comparable across datasets, and consistent with minimal symmetry and finite-resolution robustness. These requirements leave two conserved quantities: Shannon entropy $H$, which measures global uncertainty [7], and Fisher information, which measures local geometry [8]. Together, they specify the conserved informational budget of belief updating, independently of the algorithm, dynamics, or physical substrate that realises inference.

This conserved budget also exposes what remains unconstrained. Through the entropy–Fisher relation [9], entropy fixes the minimum Fisher information compatible with a given degree of global uncertainty. The residual Fisher geometry beyond this entropic baseline is captured by the information potential Φ, a non-additive, scale-invariant state function measured in nats. Φ is non-negative, vanishes uniquely for isotropic Gaussian structure, and quantifies the excess geometry associated with anisotropy, multimodality, and non-convexity.

The resulting canonical state space, ($H$, Φ), separates global uncertainty from residual geometry. Information mechanics has two regimes. Conservation concerns belief updating within an internal system, where additive projections of surprisal obey exact identities. Assimilation concerns the relation between internal representations and hidden external states, where assimilation inequalities constrain information loss and residual conditional geometry. This separation distinguishes internal coherence under a generative model from faithful representation of the world. The results below develop these regimes in turn, showing how information conservation and assimilation constrain uncertainty reduction, posterior geometry, inference difficulty, and external-state representation, without prescribing how inference is implemented.

# Results

## Information conservation and its geometric projections

Let $x \in \mathcal{X} \equiv \mathbb{R}^N$ denote latent states with prior density $p_0(x)$. After observing sensory data $o$, beliefs are updated to a posterior $p(x|o)$. Prior and posterior surprisals are defined as $S_0(x) = -\log p_0(x)$ and $S(x) = -\log p(x|o)$. Introducing the pointwise information gain $i(x) = \log\big(p(o|x)/p(o)\big)$ via Bayes' theorem [6] yields the exact identity,

$$S(x) + i(x) = S_0(x). \tag{1}$$

This expresses pointwise information conservation, measured in nats, in which any reduction in surprisal is exactly balanced by information gain (**Fig. 1a**). This relation precedes optimisation and follows directly from the reweighting of probability measures.

Because the pointwise identity holds for all $x$, any projection that preserves it must be additive, and hence linear, in the underlying information fields (see **SI.1** for details). To obtain quantities comparable across datasets, attention is therefore restricted to expectation-valued linear statistics constructed by applying a linear operator $\mathcal{L}$ to the surprisal field $S(x)$ and averaging under the posterior measure. The same construction applies to $S_0(x)$ and $i(x)$, generating a formally infinite family of projected conservation identities. Requiring translation invariance—so that the statistic does not depend on an arbitrary choice of origin—further constrains $\mathcal{L}$ to be a convolution operator, enabling a complete spectral characterisation of all admissible projections (**Fig. 1b**).

Treating surprisal as a scalar field, any linear translation-invariant statistic admits an exact representation in Fourier space, where convolution reduces to multiplication by a transfer function $\hat{\mathcal{L}}(k)$. Imposing gauge symmetry—specifically rotational invariance in the embedding space—restricts admissible filters to isotropic functions of spatial frequency, $\hat{\mathcal{L}}(k) = g(|k|^2)$. Expanding $g$ in powers of $|k|^2$ provides a natural organisation of these projections by scale, with each order corresponding to a distinct degree of geometric sensitivity, $\hat{\mathcal{L}}_m(k) = |k|^{2m}$. This construction defines a hierarchy of symmetry-consistent statistics, ranging from global to increasingly local structure (see **SI.1**).

The lowest, zero-order projection ($m = 0$) assigns equal weight to all Fourier modes and therefore captures the global component of surprisal. This projection recovers the Shannon entropy (conditional entropy) [7], $H[p] = \langle S(x) \rangle$, where $p = p(x|o)$ denotes the posterior and $\langle \cdot \rangle$ denotes the expectation under $p(x, o) = p(x|o)p(o)$ throughout the paper. Averaging equation (1) then yields the familiar conservation relation,

$$H[p] + I[o; x] = H[p_0], \tag{2}$$

where $H[p_0] = \langle S_0(x) \rangle$ denotes prior entropy and $I[o; x] = \langle i(x) \rangle$ is the mutual information. Inference therefore redistributes uncertainty from entropy into mutual information.

The next, first-order projection ($m = 1$) weights Fourier modes quadratically and therefore captures local geometric structure. This recovers the trace of the Fisher information matrix of the posterior density [8], $J[p] = \operatorname{tr} \mathcal{J}[p] = \langle \Delta S(x) \rangle$, where $\Delta$ is the Laplacian. Projecting equation (1) accordingly yields a complementary conservation,

$$J[p] - J[o; x] = J[p_0], \tag{3}$$

where $J[p_0] = \operatorname{tr} \mathcal{J}[p_0] = \langle \Delta S_0 \rangle$ denotes the prior Fisher information and $J[o; x] = -\langle \Delta i(x) \rangle = \langle |\nabla i(x)|^2 \rangle \geq 0$ is the directional mutual Fisher information, namely Fisher divergence between $p(x|o)$ and $p_0(x)$ (see **SI.2**). While entropy quantifies global uncertainty, Fisher information captures local sharpness and curvature.

Although higher-order projections ($m \geq 2$) can be formally defined, they do not yield robust state coordinates at finite resolution. The $m$-th projection weights spatial frequencies as $|k|^{2m}$, so higher orders increasingly amplify high-frequency components beyond the resolution cutoff. Across a broad class of high-frequency fluctuations of the surprisal gradient, ranging from white and pink noise to Brownian and moderately faster-decaying spectra, the first higher-order projection beyond Fisher information ($m = 2$) remains cutoff-sensitive (see **SI.3**). Finite-resolution stability therefore leaves no generic additive state coordinate beyond the two projections identified above.

In summary, although pointwise information conservation admits infinitely many formal projections, the combined requirements of additivity, symmetry, and robustness collapse them into only two distinct conserved quantities: entropy, governing global uncertainty, and Fisher information, capturing complementary local geometry. These constraints delimit how information can be redistributed without prescribing the mechanisms that realise inference.

## A scale-invariant state function for informational structure

A complete information-mechanical description requires a state function that characterises informational structure independently of coordinate scale, while remaining compatible with the conserved quantities identified above. Additive statistics that preserve the pointwise identity necessarily depend on the choice of scale, and therefore cannot by themselves capture scale-invariant information content. What is required instead is a non-additive state function constructed from the conserved global and local components, yet invariant under reparametrisation.

Shannon entropy alone is insufficient, as it increases monotonically under coarse-graining and is insensitive to residual structure. Fisher information, by contrast, captures local sharpness, but its minimum is constrained by entropy and it is intrinsically scale dependent, precluding a coordinate-scale-invariant notion of information content on its own. Taken separately, neither quantity provides a scale-invariant characterisation of informational structure. Together, however, they admit a unique non-additive state function that resolves this tension.

Crucially, requiring scale invariance fixes the functional form of such a state function (see **SI.4**). Accordingly, an information potential is defined as

$$\Phi(H, J, N) = H + \frac{N}{2} \log \frac{J}{2\pi e N}, \tag{4}$$

where $H = H[p]$ denotes the posterior entropy and $J = \mathrm{tr}\, \mathcal{I}[p]$ is the trace of the posterior Fisher information. This form is not ad hoc. Once additivity is abandoned, invariance under coordinate rescaling determines the logarithmic embedding of Fisher information, yielding a scale-invariant state function measured in nats.

The construction is further fixed by the entropy–Fisher relation. Stam's entropy power inequality [9] provides $V[p]J[p] \geq N$, where $V[p] = (2\pi e)^{-1}\exp(2H[p]/N)$ is the entropy power, equal to the variance of an isotropic Gaussian with the same entropy as $p$. Equality holds if and only if $p$ is an isotropic Gaussian. Thus, for a given entropy, Fisher information has the lower bound $J_{\min} = N/V[p]$. The information potential can therefore be written as $\Phi = (N/2)\log(J/J_{\min})$. It is non-negative, $\Phi \geq 0$, vanishes uniquely for isotropic Gaussian structure, and measures, in nats, the Fisher excess beyond the entropy-determined baseline.

Crucially, this identifies the canonical state coordinates (**Fig. 1c**). The primitive conserved projections yield $(H, J)$, yet part of $J$ is already determined by the uncertainty volume specified by $H$. The independent residual coordinate is $\Phi$. Thus, the $(H, \Phi)$ coordinates provide the minimal state space for informational structure: $H$ specifies the effective volume of plausible states, which typically decreases as evidence accumulates, while $\Phi$ quantifies the irreducible geometric structure within that volume. This residual geometry includes anisotropy, multimodality, ruggedness, and sharp transitions, capturing local structure that cannot be eliminated by naive uncertainty reduction alone. Posterior beliefs therefore admit an algorithm-independent two-coordinate description that specifies volume and scale-invariant geometry.

## Information potential asymptotically counts effective local optima

Inference and learning can be cast as the problem of identifying states that minimise surprisal $S(x)$ [4,5]. In non-convex landscapes, they can be limited by the proliferation of local minima [10].

The number of distinct basins of attraction therefore provides a natural measure of landscape complexity.

The information potential $\Phi$ can be connected to such landscape complexity. To this end, posterior structure is assessed at a finite resolution reflecting limited computational and representational resources, under which modes separated below this scale are operationally indistinguishable. This renders the effective number of distinct solutions well defined.

In the low-temperature regime, where the posterior decomposes into a mixture of well-separated Gaussian modes, $\Phi$ admits a transparent asymptotic form (see **SI.5**). Structural details within individual modes are asymptotically suppressed, leaving only their multiplicity as a relevant degree of freedom. In this limit, $\Phi$ reduces asymptotically to

$$\Phi \approx \log N_{\mathrm{LM}}\,, \tag{5}$$

to leading order, where $N_{\mathrm{LM}}$ denotes the effective number of local minima of $S(x)$ that carry non-negligible posterior weight. Hence, $\Phi$ directly quantifies the proliferation of spurious solutions that obstruct convergence to the global minimum. Numerical analyses support the asymptotic scaling predicted by equation (5) (**Fig. 1d**).

This establishes a direct link between information and computation. Exhaustive search reduces uncertainty by explicitly eliminating candidate states, with a cost set by the resolution- and dimensionality-dependent size of the admissible solution space. By contrast, information acquisition may contract the solution landscape itself. At fixed entropy, a reduction in Fisher information lowers $\Phi$ in proportion to dimensionality, thereby inducing an exponential decrease in the effective number of local minima and substantially reducing the search burden associated with competing solutions.

To further expose the structural role of $\Phi$, Gaussian coarse-graining can be viewed as the fixed-scale Gaussian reference case of information assimilation. When internal states are coupled to an isotropic Gaussian source, the induced predictive distribution follows a confined diffusion that suppresses high-frequency structure and relaxes towards the finite-variance isotropic Gaussian. Along this flow, entropy increases and Fisher information decreases, both monotonically. These changes combine into a monotone dissipation of the information potential, $\mathrm{d}\Phi/\mathrm{d}t \leq 0$, with equality only for isotropic Gaussian structure (see **SI.6**). Numerical simulations confirm this property (**Fig. 1e**). Thus, the same state function that asymptotically counts effective local optima also dissipates when those optima are blurred at fixed scale.

In this sense, information and computation can be read as interchangeable currencies. Computation compensates for missing information, but at far greater expense, whereas informative observations reshape the landscape itself. Efficient inference therefore arises from the acquisition of information that reshapes the landscape of possible solutions. Consistently, Gaussian coarse-graining dissipates the structural excess quantified by $\Phi$, driving posterior structure towards the isotropic Gaussian as the unique asymptotic fixed point of this second-law-like flow.

## Assimilation inequalities and external-state inference

Bayesian inversion is often taken to imply inference about hidden causes in the external world [4,5]. Formally, however, Bayesian and variational inversion only specify how sensory information updates posterior beliefs under an assumed generative model [11]. Exact or variational inversion therefore guarantees internal coherence, but not by itself faithful inference of environmental states. The same observations may support many internally consistent latent explanations [12,13].

Thus, Bayesian inference is, in general, inference within a model rather than inference of the world.

This exposes a basic law-level question. If internal states are defined relative to an assumed model, what can be guaranteed about their relation to the hidden states that generated sensory inputs? Assimilation inequalities address this question independently of any architecture, learning rule, or optimisation scheme. Consider the Markov chain $s \to o \to x \to o'$, in which hidden environmental states $s$ influence internal representations $x$ only through observations $o$, and $o'$ denotes reconstructed observations (**Fig. 1f**). The data processing inequality [14] provides $I[s;o] \geq I[s;x] \geq I[s;o']$, ensuring that internal states cannot contain more information about hidden causes than is already present in observations. Sharper bounds follow from the chain rule and conditional mutual information (see **SI.7**),

$$\begin{cases} H[s|o] \leq H[s|x] \leq H[s|o] + H[o|x] - H_o^* \\ H[x|o] \leq H[x|s] \leq H[x|o] + H[o|s] - H_o^* \end{cases} \tag{6}$$

where $H_o^*$ denote the irreducible entropy floor of the observation space, such that $H[o|s,x] \geq H_o^*$ for all admissible representations.

These inequalities separate intrinsic ambiguity from representational loss. The terms $H[s|o]$ and $H[o|s]$ are imposed by the world; they quantify ambiguities of external-state inference from observations and observation generation from hidden states, respectively. By contrast, $H[x|o]$ and $H[o|x]$ quantify losses introduced by internal representation and reconstruction, respectively. Hence, if reconstruction is nearly lossless, $x$ cannot be much worse than $o$ as a representation of $s$. In the ideal case $H[o|x] = H_o^*$, the externally available information is preserved and $H[s|x] = H[s|o]$. Because $o'$ is generated from $x$, conditioning on $x$ cannot leave greater uncertainty about $o$ than conditioning on its reconstruction, $H[o|x] \leq H[o|o']$. Thus, the first upper bounds can be relaxed by replacing $H[o|x]$ with the directly measurable $H[o|o']$. In essence, internal dynamics can redistribute information about hidden causes, but cannot create it.

The same Markov structure also constrains posterior local geometry. Let $J[s|x]$ denote the conditional Fisher information of the posterior over $s$ given $x$. Conditional Jensen's inequality implies that posterior sharpness about upstream causes cannot increase downstream, $J[s|x] \leq J[s|o]$ [15]. Thus, a representation may be sufficient in a Shannon sense yet geometrically poor if it retains unnecessary ruggedness.

These entropic and geometric constraints are combined by the conditional information potential, $\Phi[s|x] = H[s|x] + \frac{N_s}{2}\log\frac{J[s|x]}{2\pi e N_s}$, with expectations taken under $p(s,x)$, as well as an analogous definition for $\Phi[x|s]$. Because $\Phi \geq 0$, with equality only for isotropic Gaussian structure, $\Phi$ quantifies residual posterior complexity beyond entropy alone. Under the same Markov assumptions (see **SI.8**),

$$\begin{cases} 0 \leq \Phi[s|x] \leq \Phi[s|o] + H[o|x] - H_o^* \\ 0 \leq \Phi[x|s] \leq \Phi[x|o] + H[o|s] - H_o^* \end{cases} \tag{7}$$

These bounds strengthen ordinary sufficiency. The first states that the structural complexity of the posterior induced by the internal state, $p(s|x)$, is bounded by the sum of the intrinsic difficulty of the external inference problem, $\Phi[s|o]$, and reconstruction loss, $H[o|x]$. Thus, if the external problem is already simple and reconstruction loss is small, the posterior induced by $x$ must remain structurally simple. The second gives the complementary constraint; the complexity of $p(x|s)$ is bounded by the intrinsic complexity of the encoding kernel, $\Phi[x|o]$, and the ambiguity of observation generation, $H[o|s]$. A good representation therefore preserves information about hidden causes while maintaining simple conditional structure in both directions of the $s \leftrightarrow x$ relation.

This bidirectional constraint is essential because information preservation alone does not imply linear mappings [12,13]. A law-level route to linearity requires both conditional geometries to be constrained. In the ideal case $\Phi[s|x] = \Phi[x|s] = 0$, both conditional distributions attain Gaussian-minimal structure. Under the stated regularity assumptions, the joint *s*–*x* surprisal is then quadratic, forcing the relation between external states and internal representations to be affine (see **SI.9**). The approximate case provides a quantitative rigidity statement. Small bidirectional information potential implies near-quadratic joint surprisal and hence an almost-affine mapping, with deviations from the affine limit of order $\sqrt{\Phi[s|x] + \Phi[x|s]}$ (see **SI.10**). This scaling is numerically illustrated using multi-layer neural networks with varying nonlinearity (**Fig. 1g**).

In essence, assimilation inequalities clarify the claims licensed by Bayesian inversion. Inversion ensures coherent belief updating under a generative model. The inequalities identify the additional conditions under which internal states retain, up to controlled loss, both information about hidden causes and the geometric tractability of the induced posterior. They thereby specify the extent to which external-state inference is possible in principle, independently of architecture, optimisation scheme or learning rule.

## Discussion

Information mechanics identifies law-level constraints on inference that are independent of algorithms, dynamics, and optimisation schemes. Belief updating converts prior uncertainty into posterior uncertainty and data-dependent information gain. Under additivity, symmetry, and finite-resolution robustness, this redistribution admits two conserved projections: Shannon entropy, which constrains global uncertainty, and Fisher information, which constrains local geometric sharpness. These identities specify the informational budget within which any neural, physical, or algorithmic realisation of inference must operate.

The information potential $\Phi$ characterises the residual posterior structure left open by these conservation relations. As a scale-invariant state function, $\Phi$ vanishes only for isotropic Gaussian beliefs and increases with anisotropy, multimodality, and ruggedness. At finite resolution, $\Phi$ asymptotically reduces to the logarithm of the effective number of local minima, thereby linking posterior geometry to search difficulty. Its minimisation marks the emergence of Gaussian, effectively convex representations in high-dimensional regimes, providing an invariant account of asymptotic tractability and aligning with results on asymptotic linearisation in non-linear inference [16] and prediction [17].

Assimilation inequalities extend this framework to external-state inference. Bayesian inversion secures only internal consistency under a generative model; faithful inference about hidden environmental states requires additional constraints on what survives the passage $s \to o \to x$. Entropy inequalities quantify information loss about external causes, whereas $\Phi$-based inequalities quantify the geometric simplicity of the induced conditional posteriors. Representational quality therefore requires both Shannon-theoretic sufficiency and Fisher/$\Phi$-theoretic simplicity.

This criterion also clarifies generalisation. A complex *s*–*x* relation may support inference within the training domain while limiting extrapolation and transfer. In the Gaussian-minimal limit, bidirectional conditional simplicity restricts the admissible relation towards an affine map, supplying a transferable coordinate system for inference. Together, these results establish information mechanics as a law-level account of how informational structure constrains inference difficulty.

## References

1. Linsker, R. Self-organization in a perceptual network. *Computer* **21**, 105–117 (1988).

2. Dayan, P., Hinton, G. E., Neal, R. M. & Zemel, R. S. The Helmholtz machine. *Neural Comput.* 7, 889–904 (1995).

3. Hinton, G. E. & Salakhutdinov, R. R. Reducing the dimensionality of data with neural networks. *Science* **313**, 504–507 (2006).

4. Friston, K. J., Kilner, J. & Harrison, L. A free energy principle for the brain. *J. Physiol. Paris* **100**, 70–87 (2006).

5. Friston, K. J. The free-energy principle: a unified brain theory?. *Nat. Rev. Neurosci.* **11**, 127–138 (2010).

6. Bayes, T. LII. An essay towards solving a problem in the doctrine of chances. By the late Rev. Mr. Bayes, FRS communicated by Mr. Price, in a letter to John Canton, AMFR S. *Philos. Trans. R. Soc. Lond.* **53**, 370–418 (1763).

7. Shannon, C. E. A mathematical theory of communication. *Bell Syst. Tech. J.* **27**, 379–423 (1948).

8. Fisher, R. A. On the mathematical foundations of theoretical statistics. *Philos. Trans. A Math. Phys. Eng. Sci.* **222**, 309–368 (1922).

9. Stam, A. J. Some inequalities satisfied by the quantities of information of Fisher and Shannon. *Inf. Comput.* **2**, 101–112 (1959).

10. Jain, P. & Kar, P. Non-convex optimization for machine learning. *Found. Trends Mach. Learn.* **10**, 142-336 (2017).

11. Blei, D. M., Kucukelbir, A. & McAuliffe, J. D. Variational inference: a review for statisticians. *J. Am. Stat. Assoc.* **112**, 859–877 (2017).

12. Hyvärinen, A. & Pajunen, P. Nonlinear independent component analysis: Existence and uniqueness results. *Neural Netw.* **12**, 429–439 (1999).

13. Jutten, C. & Karhunen, J. Advances in blind source separation (BSS) and independent component analysis (ICA) for nonlinear mixtures. *Int. J. Neural Syst.* **14**, 267–292 (2004).

14. Cover, T. M. & Thomas, J. A. *Elements of informaRon theory.* John Wiley & Sons (1991).

15. Zamir, R. A proof of the Fisher information inequality via a data processing argument. *IEEE Trans. Inf. Theory* **44**, 1246–1250 (1998).

16. Isomura, T. & Toyoizumi, T. On the achievability of blind source separation for high-dimensional nonlinear source mixtures. *Neural Comput.* **33**, 1433–1468 (2021).

17. Isomura, T. & Toyoizumi, T. Dimensionality reduction to maximize prediction generalization capability. *Nat. Mach. Intell.* **3**, 434–446 (2021).

**Acknowledgements**

T.I. is supported by the Japan Society for the Promotion of Science (JSPS) KAKENHI under Grant Number JP23H04973, the Japan Agency for Medical Research and Development (AMED) under Grant Number JP23wm0625001, and the Japan Science and Technology Agency (JST) CREST under Grant Number JPMJCR22P1. The funders had no role in study design, data collection and analysis, decision to publish, or preparation of the manuscript.

**Competing interest declaration**

The author declares no competing interests.

## Figure legends

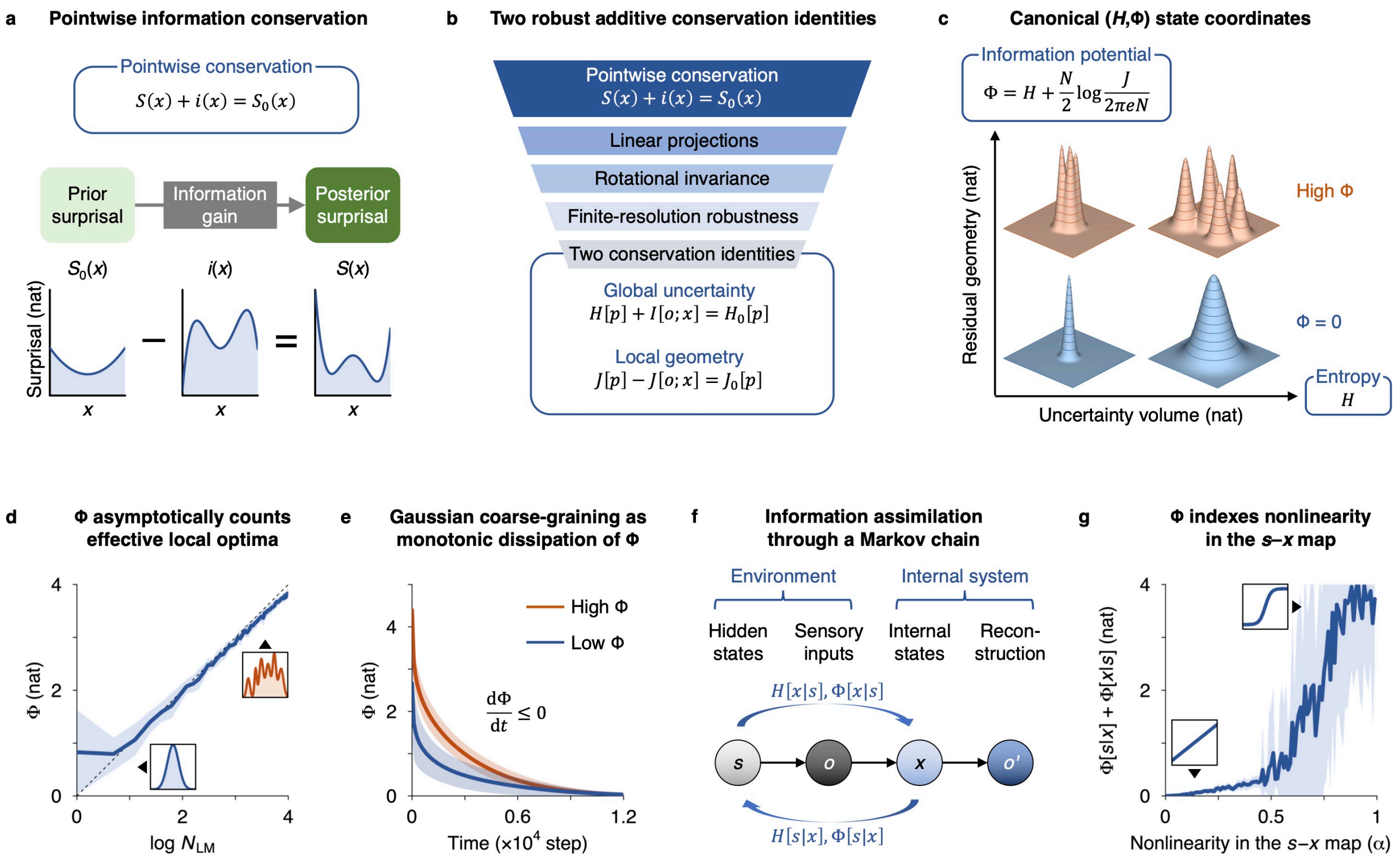

**Fig. 1. Information conservation, state coordinates, and assimilation. a,** Pointwise information conservation. Bayes' rule provides an exact identity between prior surprisal $S_0(x)$, posterior surprisal $S(x)$, and pointwise information gain $i(x)$. **b,** Additive projections of the pointwise identity. Requiring linearity, rotational invariance, and finite-resolution robustness selects two robust conservation identities for Shannon entropy $H$ and Fisher information $J$. **c,** Canonical $(H, \Phi)$ state coordinates. The information potential $\Phi$ extracts residual Fisher geometry beyond the entropy-determined baseline. Schematic densities illustrate that $H$ specifies uncertainty volume, whereas $\Phi$ increases with anisotropy, multimodality, and sharp geometric features. **d,** Relation between $\Phi$ and the effective number of local optima, $N_{\mathrm{LM}}$. In finite-resolution multimodal landscapes, $\Phi$ scales asymptotically with $\log N_{\mathrm{LM}}$. Insets show examples of low- and high-multiplicity landscapes. **e,** Gaussian coarse-graining as a reference dissipative flow. Under Gaussian smoothing, high-frequency structure is progressively blurred, entropy increases and $\Phi$ decreases monotonically. **f,** Information assimilation through a Markov chain. Hidden environmental states $s$

generate sensory observations $o$, which induce internal states $x$ and reconstructions $o'$. Assimilation inequalities constrain both entropy and information potential in the conditional relations $s|x$ and $x|s$, thereby bounding information loss and residual conditional geometry. **g,** Bidirectional information potential constrains nonlinearity in the $s$–$x$ map. Increasing nonlinearity increases $\Phi[s|x] + \Phi[x|s]$, indicating departure from the Gaussian-minimal, near-affine limit. In (d), (e), and (f), lines and shaded areas indicate means ± standard deviations. Simulations were repeated 20 times for each point using randomly generated distributions. In (d), two-dimensional distributions were generated as Gaussian mixture models with varying numbers of peaks; $H, J$, $\Phi$, and the number of local optima were then computed. In (e), initial distributions were two-dimensional Gaussian mixtures, and Fokker–Planck dynamics were used to evolve them towards an isotropic Gaussian. In (g), maps from $s$ to $x$ were defined by three-layer neural networks with random orthogonal weights. Nonlinearity of activation function $f(x) = (1 - \alpha)x + \alpha \tanh(4x)$ was controlled by varying $0 \leq \alpha \leq 1$. The conditional distribution $p(x|s)$ was modelled as a small-noise fluctuation around the network map. The networks were then inverted to induce $x$-to-$s$ maps.

# Supplementary Information on

# "Information mechanics: conservation and assimilation"

Takuya Isomura[1,2]

[1] Brain Intelligence Theory Unit, RIKEN Center for Brain Science, 2-1 Hirosawa, Wako, Saitama 351-0198, Japan

[2] RIKEN Quantum, 2-1 Hirosawa, Wako, Saitama 351- 0198, Japan

Corresponding author email: takuya.isomura@riken.jp

This file includes:

Supplementary Information SI.1 – SI.10

## SI.1 Admissible projections of pointwise information conservation

The posterior density $p(x|o)$ is assumed to have full support on $\mathcal{X} \equiv \mathbb{R}^N$, to be twice continuously differentiable, and to decay sufficiently rapidly at infinity, ensuring that boundary terms vanish and that integration by parts is well defined throughout.

Bayes' theorem [1] induces the pointwise identity in equation (1), which holds for every $x \in \mathcal{X}$. Because this identity holds pointwise, any projection that preserves it must be *additive*. Compatibility with scalar multiplication therefore restricts admissible projections to linear functionals of the underlying information fields.

To interpret such projections as statistical quantities, attention is restricted to *expectation-valued* linear functionals. Let $\mathcal{L}$ denote a linear operator acting on functions of $x$. The associated scalar statistics are defined as $\langle(\mathcal{L}S)(x)\rangle$, where $\langle\cdot\rangle = \int \cdot\, p(x,o)dxdo$ denotes expectation with respect to a joint posterior measure $p(x,o) = p(x|o)p(o)$ throughout the paper. This construction separates the choice of information pooling, encoded in $\mathcal{L}$, from the probabilistic averaging. Hence, any such operator yields a projected conservation identity,

$$\langle(\mathcal{L}S)(x)\rangle + \langle(\mathcal{L}i)(x)\rangle = \langle(\mathcal{L}S_0)(x)\rangle. \tag{8}$$

Varying $\mathcal{L}$ generates a formally infinite family of conservation identities. The objective is to identify those projections that are mutually independent, symmetry-consistent, and robust under finite resolution or noisy observations.

A key structural constraint is translation invariance. Requiring invariance under shifts, $\mathcal{L}(T_\mu f) = T_\mu(\mathcal{L}f)$ for all $\mu$, where $T_\mu f(x) = f(x-\mu)$, constrains admissible operators to commute with translations. Under standard regularity conditions, any such linear operator admits a convolution representation,

$$(\mathcal{L}S)(x) = \int K(x-y)\, S(y)\, dy, \tag{9}$$

for some kernel $K$. Consequently, translation-invariant expectation-valued linear statistics are in one-to-one correspondence with convolution operators, providing a complete and tractable characterisation of admissible projections.

Once admissible operators are restricted to convolutions, further analysis is most naturally carried out in Fourier space, where convolution reduces to multiplication. This representation renders the classification of admissible projections both explicit and tractable.

Assume $x \in \mathbb{R}^N$ with sufficient decay at infinity, or $x \in \mathbb{T}^N$ with periodic boundary conditions. The surprisal admits the exact Fourier representation

$$S(x) = \int \hat{S}(k) e^{\mathbf{i}k\cdot x} dk \,, \tag{10}$$

with inverse relation

$$\hat{S}(k) = \frac{1}{(2\pi)^{N_x}} \int S(x) e^{-\mathbf{i}k\cdot x} dx \,, \tag{11}$$

which constitutes a complete orthogonal basis. The same construction applies to $S_0(x)$ and $i(x)$.

For any translation-invariant linear operator $\mathcal{L}$ with convolution kernel $K$, the action on $S$ becomes diagonal in Fourier space,

$$\widehat{(\mathcal{L}S)}(k) = \hat{\mathcal{L}}(k)\hat{S}(k), \tag{12}$$

where $\hat{\mathcal{L}}(k) = \widehat{K}(k)$ is the spectral transfer function. Consequently, any translation-invariant expectation-valued statistic is uniquely specified by its spectral filter $\hat{\mathcal{L}}(k)$.

A further and crucial restriction follows from gauge symmetry. Gauge symmetry here refers simply to rotational invariance in the embedding space. That is, requiring invariance under rotations $k \mapsto Rk$ with $R \in SO(N)$ ensures that admissible projections do not depend on an arbitrary choice of coordinate orientation. In Fourier space, this requirement constrains the spectral transfer function to satisfy $\hat{\mathcal{L}}(Rk) = \hat{\mathcal{L}}(k)$ for all $R \in SO(N)$. For invariance to hold for all rotations, the transfer function can depend only on rotationally invariant quantities. The only such scalar built from $k$ is its squared magnitude, $|k|^2$. Admissible filters therefore reduce to radial functions of the form

$$\hat{\mathcal{L}}(k) = g(|k|^2), \tag{13}$$

for some scalar function $g$. Expanding $g$ as a polynomial,

$$g(|k|^2) = \sum_{m=0}^{M} a_m |k|^{2m} \tag{14}$$

yields, in real space, a hierarchy of differential operators generated by successive powers of the Laplacian, $\Delta$.

Therefore, before robustness is imposed, the distinct orders of the isotropic expansion are indexed by the monomials $g_m(|k|^2) = |k|^{2m}$. For those orders $m$ for which the derivatives and corresponding expectations are well defined, these monomials define

$$(\widehat{\mathcal{L}_m S})(k) = |k|^{2m}\hat{S}(k) \Leftrightarrow (\mathcal{L}_m S)(x) = (-\Delta)^m S(x). \tag{15}$$

The corresponding statistics take the form $\langle(\mathcal{L}_m S)(x)\rangle = \langle(-\Delta)^m S(x)\rangle$, providing successive orders of isotropic geometric information. A radial filter admitting the expansion in equation (14) is a superposition of these orders and does not constitute an additional order of the expansion.

Taken together, admissible projections of pointwise information conservation form a strict hierarchy: additive maps ⊃ linear functionals ⊃ expectation-valued linear statistics ⊃ translation-invariant (convolution) operators ⊃ isotropic (rotational symmetry) spectral filters $g(|k|^2)$. This hierarchy delineates progressively stronger structural constraints and provides the foundation for the robustness analysis and the subsequent reduction to low-order invariants used in the main text.

## SI.2 Derivation of mutual Fisher information

Mutual Fisher information defined by $J[o; x] \coloneqq -\langle \Delta i \rangle$ can also be written in gradient form. Unless stated otherwise, $\langle \cdot \rangle = \int \cdot\ p(x, o)\, dx\, do$. The squared gradient norm is expressed as $\langle |\nabla i|^2 \rangle = \langle |\nabla S - \nabla S_0|^2 \rangle = \langle |\nabla S|^2 \rangle - 2\langle \nabla S \cdot \nabla S_0 \rangle + \langle |\nabla S_0|^2 \rangle$. From integration by parts, the cross term satisfies

$$\langle \nabla S \cdot \nabla S_0 \rangle = \int \nabla S \cdot \nabla S_0\, p\, dx = -\int \nabla S_0 \cdot \nabla p\, dx = \int \Delta S_0\, p\, dx = \langle \Delta S_0 \rangle. \tag{16}$$

Here, $J[p_0] = \langle \Delta S_0 \rangle = \langle |\nabla S_0|^2 \rangle$ as $p_0$ is independent of $o$ and thus $\int \Delta S_0 p(x, o) dx do = \int \Delta S_0 p_0(x) dx$. Therefore, $\langle |\nabla i|^2 \rangle = \langle |\nabla S|^2 \rangle - \langle \Delta S_0 \rangle$. Using the Fisher conservation relation in equation (3), together with $\langle |\nabla S|^2 \rangle = \langle \Delta S \rangle$, it follows

$$J[o; x] = -\langle \Delta i \rangle = \langle |\nabla i|^2 \rangle \geq 0. \tag{17}$$

Thus, the posterior Fisher information satisfies $J[p] \geq J[p_0]$.

## SI.3 Finite-resolution robustness excludes higher-order projections

This section formalises how finite resolution and noise collapse the infinite family of symmetry-consistent projections to two robust conservation identities. *Robustness* is defined as stability of the projected statistic under a finite spectral cutoff: the statistic must not be dominated by arbitrarily high-frequency components whose content is noise-dependent.

Let $\mathcal{L}$ be a translation-invariant isotropic linear operator with spectral transfer function $\hat{\mathcal{L}}(k) = g(|k|^2)$. If Fourier modes above a cutoff $k_c$ are noise-dominated, then filters with rapidly increasing $g$ amplify the cutoff region and become cutoff-dependent. Robustness therefore motivates restricting $g$ to its lowest non-trivial orders, $g(|k|^2) = a_0 + a_1|k|^2$. At this stage, two symmetry-allowed candidates remain: the constant mode ($a_0 = 1, a_1 = 0$) and the quadratic mode ($a_0 = 0, a_1 = -1$). Using $\widehat{\Delta f}(k) = -|k|^2 \hat{f}(k)$, these modes correspond in real space to $\mathcal{L}_0 = 1$ and $\mathcal{L}_1 = \Delta$, yielding the associated conserved projections, $H = \langle \mathcal{L}_0 S \rangle$ and $J = \langle \mathcal{L}_1 S \rangle$, in equations (2) and (3).

Having said this, it remains to be justified that higher-order polynomial filters $\hat{\mathcal{L}}_m(k) = g_m(|k|^2) \propto |k|^{2m}$ with $m \geq 2$ are generically non-robust. To assess robustness, the posterior expectation $\langle \mathcal{L} S \rangle$ is analysed at leading order around an isotropic Gaussian posterior. At this order, both the posterior density and the surprisal decompose additively across dimensions, and cross-derivative terms in $\Delta^m$ do not contribute. Consequently, the multidimensional problem reduces to an effective one-dimensional evaluation, with non-separable corrections entering only at subleading order and neglected here.

Under this reduction to an effective one-dimensional problem, the posterior expectation can be evaluated in Fourier space using the standard identity,

$$\langle \mathcal{L} S \rangle = \int p(x|o)(\mathcal{L} S)(x)dx = \frac{1}{2\pi} \int \hat{p}(-k)\hat{\mathcal{L}}(k)\hat{S}(k)dk\,. \tag{18}$$

To assess robustness without imposing assumptions on structured low-frequency components, consider the cumulative variance of the filtered statistic restricted to the noise-dominated tail,

$$V_m(k) = \mathrm{Var}\left[\frac{1}{2\pi} \int_{|\kappa| \geq k} \hat{p}\,(-\kappa)\hat{\mathcal{L}}(\kappa)\hat{S}(\kappa)d\kappa\right]. \tag{19}$$

For isotropic operators in the monomial family, $\hat{\mathcal{L}}(k) = |k|^{2m}$ with $m \geq 0$.

To assess robustness, it is sufficient to characterise the high-frequency roughness of the surprisal field. Because roughness concerns local spatial variations, it is naturally characterised through spatial increments, represented here by the surprisal gradient $\nabla S$. Consider the regime in which the posterior is close to a Gaussian, $p(x|o) \approx \mathcal{N}[0, \sigma^2]$, so that $|\hat{p}(k)| \propto e^{-\sigma^2|k|^2/2}$ holds approximately. Beyond a cutoff scale $k_c$, the fluctuations are modelled as independent and mean-zero in Fourier space, with gradient spectrum $\mathbb{E}\left[\left|\widehat{\nabla S}(k)\right|^2\right] \propto |k|^{-\alpha}$, where $\alpha \geq 0$. White, pink, and Brownian-type gradient spectra correspond to $\alpha$ = 0, 1, and 2, respectively, while faster-decaying gradient spectra give $\alpha > 2$. Since $\widehat{\nabla S}(k) = \mathbf{i}k\hat{S}(k)$, the corresponding surprisal spectrum is $\mathbb{E}\left[\left|\hat{S}(k)\right|^2\right] \propto |k|^{-\alpha-2}$. At leading order around an isotropic Gaussian, cross-derivative contributions are absent, and the typical magnitude of post-cutoff fluctuations is therefore controlled by their variance. Under these conditions, cross terms vanish and the cumulative variance of the filtered statistic obeys

$$V_m(k) \propto \int_{|\kappa|\geq k} |\hat{p}(\kappa)|^2 \left|\hat{\mathcal{L}}(\kappa)\right|^2 \mathbb{E}\left[\left|\hat{S}(\kappa)\right|^2\right] d\kappa \propto \int_{|\kappa|\geq k} e^{-\sigma^2|\kappa|^2} |\kappa|^{4m-\alpha-2} d\kappa\,. \tag{20}$$

The sensitivity of the tail variance to the cutoff is governed by the density

$$\rho_m(k) = -\frac{d}{dk} V_m(k) \propto e^{-\sigma^2|k|^2} |k|^{4m-\alpha-2}, \tag{21}$$

which attains its maximum at $|k^*|^2 = (4m - \alpha - 2)/(2\sigma^2)$ when $4m - \alpha - 2 \geq 0$. Taking the natural finite-resolution scale $k_c \sim 1/\sigma$, robustness requires the dominant contribution to the tail variance not to lie deep inside the noise-dominated regime, i.e., $|k^*| \lesssim k_c$. This condition implies $4m - \alpha \lesssim 4$; otherwise, the dominant contribution lies beyond the resolution cutoff. For $m = 1$, which yields Fisher information, the condition is satisfied throughout the assumed range $\alpha \geq 0$; thus, the dominant contribution does not lie beyond the resolution cutoff. For the first higher-order projection beyond Fisher information, $m = 2$, robustness would require $\alpha \gtrsim 4$. Thus, throughout the broad range $0 \leq \alpha < 4$, the $m = 2$ statistic remains sensitive to cutoff-dependent high-frequency fluctuations. This range includes white, pink, Brownian-type, and moderately faster-decaying gradient spectra. Higher orders require progressively steeper decay, $\alpha \gtrsim 4m - 4$, and are therefore still less generically robust.

Notably, choosing a different radial filter does not introduce an additional robust component. For example, the Gaussian filter $g(|k|^2) = e^{-|k|^2}$ is stable as a whole because its higher-order terms combine to suppress high-frequency growth. However, this overall stability does not render the individual orders $m \geq 2$ robust: as shown above, each remains dominated by cutoff-dependent high-frequency fluctuations. These terms may alter the finite value of the filtered statistic, but they do not provide additional robust information. Hence, the only robust components of the isotropic expansion are the $m = 0$ and $m = 1$ terms.

In summary, although higher-order polynomial filters ($m \geq 2$) in principle define additional statistics, sufficiently high spatial frequencies are typically noise dominated and do not carry robust information. As a result, such filters become increasingly sensitive to spectral cutoffs and fail to define stable invariants under finite resolution. For white or moderately decaying high-frequency spectra of the surprisal gradient, $0 \leq \alpha < 4$, the first higher-order projection beyond Fisher information is already sensitive to cutoff-dependent fluctuations, with higher orders requiring progressively faster decay. Under rotational symmetry, this leaves only two robust isotropic components: a constant mode ($m = 0$), corresponding to the global entropy $H$, and a quadratic mode ($m = 1$), corresponding to the trace of the Fisher information $J$. The quadratic term therefore provides the unique lowest-order statistic that is both sensitive to local structure and robust under coarse-graining, furnishing a non-trivial complement to entropy.

## SI.4 Uniqueness of the information potential under gauge invariance

This section establishes the information potential $\Phi$ as the *unique* non-additive state function compatible with information conservation and minimal symmetry requirements. The argument rests upon a fundamental tension between additivity and gauge invariance, and shows that resolving this tension uniquely fixes the functional form of $\Phi$.

Belief updating restricts identity-preserving aggregations to additive statistics. Under minimal symmetry and robustness assumptions, these reduce to entropy $H[p]$ and Fisher information $J[p]$. These quantities exhaust the space of additive statistics permitted by the stated symmetry and robustness. However, additivity entails an unavoidable consequence. Because such statistics are defined as expectation values of local fields, they necessarily depend on the choice of coordinate scale.

Under a rescaling of latent variables $x \mapsto ax$, entropy and Fisher information transform as $H[p_a] = H[p] + N \log a$ and $J[p_a] = J[p]/a^2$, where $N$ denotes the dimensionality of the latent space ($x \in \mathbb{R}^N$). This scale dependence is a structural consequence of linearity and expectation-valued construction. Thus, while $H$ and $J$ enter exact conservation identities under Bayesian updating, they are *not* invariant under changes of coordinate units. Any additive statistic that preserves the pointwise identity necessarily inherits this gauge dependence.

Informational content, however, should not depend on an arbitrary choice of units or coordinate scale. A mere reparametrisation of the latent space should not alter the information carried by a belief distribution. This requirement of gauge invariance directly conflicts with additivity, as all additive information-preserving statistics are scale dependent. If information conservation is to be retained as a first-principles constraint, while gauge invariance is imposed as a requirement for meaningful information content, then additivity must be abandoned. What is required is therefore a *non-additive state function* that depends only on the conserved quantities $H$ and $J$, yet remains invariant under coordinate rescaling.

Gauge invariance uniquely fixes the functional form of such a state function. Consider a scalar function $\Phi = \Phi(H, J, N)$ constructed solely from the two conserved quantities and dimensionality. Invariance under rescaling $x \mapsto ax$ requires

$$\Phi\left(H + N \log a \, , \frac{J}{a^2}, N\right) = \Phi(H, J, N) \tag{22}$$

for all $a > 0$. Differentiating with respect to $a$ yields the first-order partial differential equation,

$$N \frac{\partial \Phi}{\partial H} - 2J \frac{\partial \Phi}{\partial J} = 0. \tag{23}$$

The general solution of this equation is

$$\Phi(H,J,N) = F\left(H + \frac{N}{2} \log J \, , N\right), \tag{24}$$

where $F$ is an arbitrary monotonic function of its first argument. Hence, scale invariance fixes the invariant combination $H + (N/2) \log J$, and the remaining freedom corresponds to a scalar reparametrisation of the same state coordinate.

Requiring $\Phi$ to serve as a state function, whose differences and dissipation rates are not altered by a state-dependent nonlinear calibration, excludes nonlinear choices of $F$ and leaves only an affine calibration. The remaining multiplicative freedom is fixed by requiring $\Phi$ to be expressed on the same nat scale as entropy, yielding

$$\Phi(H,J,N) = H + \frac{N}{2} \log J + C(N), \tag{25}$$

where $C(N)$ is a constant depending only on dimensionality. Choosing the isotropic Gaussian as the reference state fixes this constant, yielding equation (4), which vanishes if and only if $p$ is an isotropic Gaussian. In essence, scale invariance, the state-function requirement, and nat-scale and nonnegative normalisations determine the form of $\Phi$.

In summary, the information potential $\Phi$ measures, in nats, the residual structural degrees of freedom left unconstrained by the two additive conservation identities. Entropy and Fisher information specify what is preserved under belief updating; $\Phi$ quantifies what is not fixed by those constraints while remaining invariant under rescaling. In this sense, $\Phi$ plays the role of a state function for informational structure.

## SI.5 Asymptotic relation between information potential and effective local minima

This section clarifies how the information potential $\Phi$ relates to the effective number of local optima in a non-convex inference landscape. Consider a Boltzmann distribution over a continuous state space $x \in \mathcal{X} \equiv \mathbb{R}^N$, $p(x|o) \propto e^{-\beta E(x)}$, where $E(x)$ denotes a smooth energy function and $\beta \gg 1$ is the inverse temperature, such that the surprisal satisfies $S(x) = \beta E(x)$. Assume that $E(x)$ admits $N_{\mathrm{LM}}$ well-separated local minima $\{x_l\}_{l=1}^{N_{\mathrm{LM}}}$. In the low-temperature regime, each minimum admits a local quadratic approximation,

$$E(x) \approx E_l + \frac{1}{2}(x - x_l)^{\mathrm{T}} \Lambda_l (x - x_l), \tag{26}$$

where $\Lambda_l = (\nabla^2 E)(x_l) \succ 0$ is the Hessian matrix at the minimum. Under this approximation, the posterior density can be expressed as a Gaussian mixture,

$$p(x|o) \approx \sum_{l=1}^{N_{\mathrm{LM}}} w_l \mathcal{N}[x_l, \Sigma_l], \tag{27}$$

with weights $w_l = \frac{e^{-\beta E_l} |\Sigma_l|^{1/2}}{\sum_{l'=1}^{N_{\mathrm{LM}}} e^{-\beta E_{l'}} |\Sigma_{l'}|^{1/2}}$.

To define an operational notion of distinct solutions, finite resolution is introduced. First, the covariance of each Gaussian component is expressed as $\Sigma_l = \delta I + \beta^{-1}\Lambda_l^{-1}$, where $0 < \delta \ll 1$ specifies the resolution of latent variables—encoded either as an isotropic baseline covariance or as an effective noise scale. Expanding in $\beta^{-1}$ yields $\Sigma_l^{-1} = \delta^{-1}(I + (\delta\beta)^{-1}\Lambda_l^{-1})^{-1} = \delta^{-1}\left(I - (\delta\beta)^{-1}\Lambda_l^{-1} + \mathcal{O}((\delta\beta)^{-2})\right)$. Second, only components satisfying $w_l/w_{\max} \geq \varepsilon$ are retained, where $0 < \varepsilon \ll 1$ defines a density resolution and $w_{\max}$ is the maximal weight. Under these criteria, $N_{\mathrm{LM}}$ denotes the effective number of local minima contributing to the posterior $S(x)$.

In the limit $\beta \gg 1$, overlaps between distinct modes are exponentially suppressed, up to the finite-resolution floor set by $\delta$, and cross-terms between mixture components become negligible. Under this separation, the entropy is decomposed as

$$H[p] \approx H(w) + \sum_{l=1}^{N_{\mathrm{LM}}} w_l \left\{\frac{1}{2}\log(2\pi e)^N |\Sigma_l|\right\}$$

$$= H(w) + \sum_{l=1}^{N_{\mathrm{LM}}} w_l \left\{\frac{N}{2}\log 2\pi e - \frac{1}{2}\log\left|\frac{1}{\delta}\left(I - (\delta\beta)^{-1}\Lambda_l^{-1} + \mathcal{O}((\delta\beta)^{-2})\right)\right|\right\}$$

$$= H(w) + \frac{N}{2}\log 2\pi e - \frac{N}{2}\overline{\log\lambda}, \tag{28}$$

where $H(w) = -\sum_{l=1}^{N_{\mathrm{LM}}} w_l \log w_l$ is the discrete entropy of the mixture weights and $\overline{\log\lambda} = -\log\delta + \frac{1}{N}\sum_{l=1}^{N_{\mathrm{LM}}} w_l \log|I - (\delta\beta)^{-1}\Lambda_l^{-1} + \mathcal{O}((\delta\beta)^{-2})|$ denotes the averaged log-eigenvalue across modes. Here, overlap corrections are neglected, as low temperature and sufficient mode separation render these contributions exponentially suppressed in $\beta$, up to the finite-resolution floor set by $\delta$, and subleading for the entropy.

Under sufficient mode separation, posterior mass concentrates within individual modes, rendering boundary contributions exponentially small. In this regime, the Fisher information is dominated by within-mode curvature,

$$J[p] \approx \mathrm{tr}\left[\sum_{l=1}^{N_{\mathrm{LM}}} w_l \Sigma_l^{-1}\right] = \sum_{l=1}^{N_{\mathrm{LM}}} w_l \mathrm{tr}\left[\frac{1}{\delta}\Big(I - (\delta\beta)^{-1}\Lambda_l^{-1} + \mathcal{O}((\delta\beta)^{-2})\Big)\right] = N\bar{\lambda}, \tag{29}$$

where $\bar{\lambda} = \frac{1}{\delta}\Big(1 - \frac{1}{N}\sum_{l=1}^{N_{\mathrm{LM}}} w_l \mathrm{tr}[\Lambda_l^{-1}]\,(\delta\beta)^{-1} + \mathcal{O}((\delta\beta)^{-2})\Big)$ denotes the averaged eigenvalue across modes. Hence, substituting equations (28) and (29) into the information potential $\Phi$ in equation (4) yields

$$\Phi \approx H(w) + \frac{N}{2}\big(\log\bar{\lambda} - \overline{\log\lambda}\big). \tag{30}$$

Expanding in powers of $(\delta\beta)^{-1}$ shows that $\overline{\log\lambda} = -\log\delta - \frac{1}{N}\sum_{l=1}^{N_{\mathrm{LM}}} w_l \mathrm{tr}[\Lambda_l^{-1}]\,(\delta\beta)^{-1} + \mathcal{O}((\delta\beta)^{-2}) = -\log\delta + \big(\delta\bar{\lambda} - 1\big) + \mathcal{O}((\delta\beta)^{-2})$ and $\log\bar{\lambda} = -\log\delta + \big(\delta\bar{\lambda} - 1\big) + \mathcal{O}((\delta\beta)^{-2})$. Thus, the curvature-dependent correction in equation (30) cancels at leading order when $|\Lambda_l^{-1}|(\delta\beta)^{-1} \ll 1$ uniformly across modes, yielding $\frac{N}{2}\big(\log\bar{\lambda} - \overline{\log\lambda}\big) = \mathcal{O}(N(\delta\beta)^{-2})$. In this regime, local curvatures become approximately isotropic and the geometric contribution to $\Phi$ is asymptotically suppressed.

The remaining term $H(w)$ controls the leading behaviour. Its maximum occurs for uniform weights $w_l = 1/N_{\mathrm{LM}}$, yielding $H(w) = \log N_{\mathrm{LM}}$. The minimum corresponds to one dominant mode with $w_l/w_{\max} = \varepsilon$ for all others. In this case, from $w_{\max} + (N_{\mathrm{LM}} - 1)\varepsilon w_{\max} = 1 \Leftrightarrow w_{\max} = 1/\big(1 + \varepsilon(N_{\mathrm{LM}} - 1)\big)$, $H(w) = -w_{\max}\log w_{\max} - (N_{\mathrm{LM}} - 1)\varepsilon w_{\max}\log\varepsilon w_{\max} = \log\big(1 + \varepsilon(N_{\mathrm{LM}} - 1)\big) - \frac{\varepsilon(N_{\mathrm{LM}}-1)}{1+\varepsilon(N_{\mathrm{LM}}-1)}\log\varepsilon$. When $\varepsilon N_{\mathrm{LM}} \gg 1$, $H(w) = \log N_{\mathrm{LM}} + \mathcal{O}\left(\frac{|\log\varepsilon|}{\varepsilon N_{\mathrm{LM}}}\right)$ to leading order. Consequently, when $N_{\mathrm{LM}} \gg |\log\varepsilon|/\varepsilon$ and $\beta \gg \sqrt{N}/\delta$, the information potential in equation (30) reduces to

$$\Phi = \log N_{\mathrm{LM}} + \mathcal{O}\left(\frac{|\log \varepsilon|}{\varepsilon N_{\mathrm{LM}}}\right) + \mathcal{O}\left(\frac{N}{(\delta\beta)^2}\right) \tag{31}$$

to the leading order.

In summary, for a Boltzmann distribution with well-separated minima, entropy and Fisher information decompose into discrete and local contributions. Under finite resolution, curvature-dependent terms asymptotically cancel. In this regime, $\Phi$ asymptotically counts distinct solutions and reduces to the logarithm of the effective number of local minima, formalising its interpretation as an information-theoretic measure of landscape complexity.

## SI.6 Monotone dissipation of the information potential under confined Gaussian coarse-graining

This section establishes the monotone decrease of the information potential under confined Gaussian coarse-graining with fixed variance $\sigma^2$. The coarse-graining is read as a confined heat flow generated by isotropic diffusion together with a linear restoring drift that preserves normalisation and the centred second moment $\langle |x-\mu|^2 \rangle = N\sigma^2$. This flow suppresses high-frequency structure and has the finite-variance isotropic Gaussian density $\mathcal{N}[\mu, \sigma^2 I]$ as its unique fixed point. The same flow admits a Bayesian interpretation as the prediction step of a dynamic Bayesian filter with isotropic Gaussian process noise and a linear mean-reverting state transition. The associated change in surprisal retains the pointwise balance form of equation (1), with $i(x)$ induced by the prediction-induced density ratio.

The Lyapunov property follows from standard entropy–Fisher identities for this confined heat flow. Entropy production obeys de Bruijn's identity [2] with the confining-drift correction, whereas Fisher information dissipates according to the Bochner identity [3] with the corresponding correction. In the information potential, these drift terms cancel, yielding $\Phi \geq 0$ and $\mathrm{d}\Phi/\mathrm{d}t \leq 0$, with equality only at the isotropic Gaussian fixed point. By translational invariance, the derivation is carried out in centred coordinates, $\mu = 0$. The result for a general mean can be recovered by translating $x \mapsto x - \mu$.

Let $p_t = p_t(x)$ denote a probability density on $\mathcal{X} \equiv \mathbb{R}^N$, and let $\langle x \rangle = 0$ and $\langle |x|^2 \rangle = N\sigma^2$ be fixed. The confined heat flow is generated by the Fokker–Planck dynamics

$$\partial_t p_t = \frac{1}{\sigma^2} \nabla \cdot (x p_t) + \Delta p_t, \tag{32}$$

assuming sufficient smoothness and decay at infinity to justify integrations by parts. This flow preserves the centred second moment, since $d\langle |x|^2 \rangle / dt = 2(N - \langle |x|^2 \rangle / \sigma^2) = 0$. Define the associated surprisal $S_t = S_t(x) = -\log p_t(x)$ and expectation $\langle \cdot \rangle = \int \cdot \, p_t(x) dx$. The Shannon entropy, Fisher information, and information potential are $H = H[p_t] = \langle S_t \rangle$, $J = J[p_t] = \langle \Delta S_t \rangle$, and $\Phi = H + (N/2) \log J + \text{const.}$

(1) Entropy production. Differentiating $H$ along the confined heat flow provides $\dot{H} = -\int \partial_t p_t \log p_t \, dx = -\int \Delta p_t \log p_t \, dx - \sigma^{-2} \int \nabla \cdot (x p_t) \log p_t \, dx$, where normalisation ensures $\int \partial_t p_t dx = 0$. Integration by parts yields $\int \Delta p_t \log p_t \, dx = -\int \nabla p_t \cdot \nabla \log p_t \, dx = -\int p_t |\nabla \log p_t|^2 dx = -J$ and $\int \nabla \cdot (x p_t) \log p_t \, dx = -\int x \cdot \nabla p_t dx = N$. Thus, de Bruijn's identity [2] with the confining correction provides

$$\dot{H} = J - \frac{N}{\sigma^2} \geq 0, \tag{33}$$

where the inequality follows from the trace Cramér–Rao inequality $J \geq N/\sigma^2$.

(2) Fisher dissipation. Differentiating Fisher information $J = \langle |\nabla S_t|^2 \rangle$ yields $\dot{J} = \int \{ (\partial_t p_t) |\nabla S_t|^2 + 2 p_t \nabla S_t \cdot \partial_t (\nabla S_t) \} dx$. The diffusive and confining components can be evaluated separately. For the heat-flow component, $\partial_t p_t = \Delta p_t$, the surprisal evolves as $\partial_t S_t = -\Delta p_t / p_t =$ $\Delta S_t - |\nabla S_t|^2$, where $\Delta p_t = \nabla \cdot \nabla p_t = -\nabla \cdot (p_t \nabla S_t) = p_t(|\nabla S_t|^2 - \Delta S_t)$ has been used. Hence, $\dot{J}_{\text{diff}} = \int \{(\Delta p_t)|\nabla S_t|^2 + 2p_t \nabla S_t \cdot \nabla(\Delta S_t - |\nabla S_t|^2)\} dx$.

Using the Bochner identity [3], $\nabla S_t \cdot \nabla(\Delta S_t) = (1/2)\Delta|\nabla S_t|^2 - |\nabla^2 S_t|_F^2$, this becomes $\dot{J}_{\text{diff}} = \int \{(\Delta p_t)|\nabla S_t|^2 + 2p_t\left((1/2)\Delta|\nabla S_t|^2 - |\nabla^2 S_t|_F^2\right) - 2p_t \nabla S_t \cdot \nabla|\nabla S_t|^2\} dx$. Integration by parts provides $\int (\Delta p_t)|\nabla S_t|^2 dx = -\int \nabla p_t \cdot \nabla|\nabla S_t|^2 dx$ and $\int p_t \Delta|\nabla S_t|^2 dx = -\int \nabla p_t \cdot \nabla|\nabla S_t|^2 dx$. Since $p_t \nabla S_t = -\nabla p_t$, the derivative terms cancel, leaving $\dot{J}_{\text{diff}} = -2\int p_t |\nabla^2 S_t|_F^2 dx = -2\langle |\nabla^2 S_t|_F^2 \rangle \le 0$.

For the confining drift, the drift component $\partial_t p_t = \sigma^{-2}\nabla \cdot (x p_t)$ provides $\partial_t S_t = -\sigma^{-2}\nabla \cdot (x p_t)/p_t = \sigma^{-2} x \cdot \nabla S_t - \sigma^{-2} N$. Substitution into the differentiated expression for $J$ yields $\dot{J}_{\text{drift}} = \sigma^{-2}\int \nabla \cdot (x p_t)|\nabla S_t|^2 dx + 2\sigma^{-2}\int p_t \nabla S_t \cdot \nabla(x \cdot \nabla S_t) dx$. The first term integrates by parts to $-2\sigma^{-2}\int p_t \nabla S_t \cdot \nabla^2 S_t x dx$, whereas the second term equals $2\sigma^{-2} J + 2\sigma^{-2}\int p_t \nabla S_t \cdot \nabla^2 S_t x dx$. The cross terms cancel, and therefore $\dot{J}_{\text{drift}} = 2\sigma^{-2} J$. Combining the diffusive and confining contributions provides the Fisher dissipation identity

$$\dot{J} = -2\langle |\nabla^2 S_t|_F^2 \rangle + \frac{2J}{\sigma^2} \le 0. \tag{34}$$

The final inequality follows from $\langle |\nabla^2 S_t|_F^2 \rangle \ge J^2/N$ and the fixed-variance Cramér–Rao bound $J \ge N/\sigma^2$.

(3) Lyapunov property of $\Phi$. From equations (33) and (34), differentiating $\Phi = H + (N/2)\log J + \text{const.}$ yields

$$\dot{\Phi} = \dot{H} + \frac{N}{2J}\dot{J} = J - \frac{N}{J}\langle |\nabla^2 S_t|_F^2 \rangle. \tag{35}$$

The confining contribution in equations (33) and (34) cancels exactly from the evolution of $\Phi$. To relate $\langle |\nabla^2 S_t|_F^2 \rangle$ to $J$, applying the pointwise inequality $|A|_F^2 \ge (\operatorname{tr} A)^2/N$, with equality if $A \propto I$, to $\nabla^2 S_t$ yields $|\nabla^2 S_t|_F^2 \ge (\Delta S_t)^2/N$. Taking expectations and applying Jensen's inequality to $\Delta S_t$ under $p_t$ provides

$$\langle |\nabla^2 S_t|_F^2 \rangle \geq \frac{\langle (\Delta S_t)^2 \rangle}{N} \geq \frac{\langle \Delta S_t \rangle^2}{N} = \frac{J^2}{N}, \tag{36}$$

Hence, substituting equation (36) into equation (35) yields

$$\dot{\Phi} \leq J - \frac{N}{J}\frac{J^2}{N} = 0. \tag{37}$$

Equality holds if and only if both inequalities in equation (36) are tight, which requires $\nabla^2 S_t$ to be proportional to the identity matrix almost everywhere and $\Delta S_t$ to be constant under $p_t$. These conditions characterise isotropic Gaussian densities. Because the confined flow preserves the centred second moment, the unique fixed point is $p_\infty(x) = \mathcal{N}[\mu, \sigma^2 I]$, for which $\Phi = 0$. Therefore, $\Phi$ is strictly decreasing under confined Gaussian coarse-graining unless $p_t$ is already the finite-variance isotropic Gaussian fixed point.

## SI.7 Derivation of entropy-based assimilation inequalities

Because hidden external states affect internal representations only through observations, the relevant variables form the Markov chain $s \to o \to x \to o'$, where $s$ denotes hidden external states, $o$ sensory inputs, $x$ internal representations, and $o'$ reconstructed inputs. This Markov structure implies the conditional independence $s \perp x | o$.

The data processing inequality provides $I[s; x] \leq I[s; o]$ [4]. Using $I[s; x] = H[s] - H[s|x]$ and $I[s; o] = H[s] - H[s|o]$, this yields

$$H[s|o] \leq H[s|x]. \tag{38}$$

Thus, encoding cannot increase information about hidden causes beyond what is already available in observations. An upper bound follows from the chain rule for conditional mutual information. Since $s \perp x | o$, one has $H[s|o, x] = H[s|o]$, and hence $H[s|x] - H[s|o] = H[s|x] - H[s|o, x] =$ $I[s; o|x]$. Let $H_o^*$ denote the irreducible entropy floor of the observation space, such that $H[o|s, x] \geq H_o^*$ for all admissible representations. Because $I[s; o|x] \leq H[o|x] - H_o^*$, it follows

$$H[s|x] \leq H[s|o] + H[o|x] - H_o^*. \tag{39}$$

Because o′ is generated from $x$, the data processing inequality for $o \to x \to o'$ provides $I[o; x] \geq I[o; o']$, or equivalently $H[o|x] \leq H[o|o']$. Combining these relations provides

$$H[s|o] \leq H[s|x] \leq H[s|o] + H[o|x] - H_o^* \leq H[s|o] + H[o|o'] - H_o^*. \tag{40}$$

This inequality separates two contributions to external-state uncertainty. The term $H[s|o]$ is fixed by the sensory ambiguity of the world, whereas $H[o|x]$ and $H[o|o']$ quantify representation or reconstruction loss. When reconstruction is nearly lossless and reconstruction uncertainty approaches the irreducible observational floor, $x$ preserves nearly all information about $s$ that is available in $o$.

A complementary bound is obtained by applying the same argument to the conditional structure of $x$. From the data processing inequality, $I[x; s] \leq I[x; o]$, and therefore $H[x|o] \leq H[x|s]$. Moreover, using $H[x|o, s] = H[x|o]$, which again follows from the Markov structure, $H[x|s] - H[x|o] = H[x|s] - H[x|o, s] = I[x; o|s]$. Because $I[x; o|s] \leq H[o|s] - H_o^*$, this gives

$$H[x|o] \leq H[x|s] \leq H[x|o] + H[o|s] - H_o^*. \tag{41}$$

Equations (40) and (41) define entropy-based assimilation inequalities. The lower bounds follow from data processing; internal states cannot contain more information about external causes than is present in observations. The upper bounds follow by bounding conditional mutual information with observational or reconstructive uncertainty. A good representation therefore compresses observations while preserving the information they carry about hidden states, so that $x$ approximates a sufficient statistic of $o$ for inferring $s$. These inequalities specify a fundamental limit on assimilation, that is, internal dynamics can redistribute information about external causes, but cannot create it.

## SI.8 Derivation of information-potential-based assimilation inequalities

The entropy inequalities in **SI.7** have a geometric counterpart. Let $s \to o \to x \to o'$ be the Markov chain defined above. For any variable $y$, define the conditional Fisher information of the posterior over $s$ by $J[s|y] \coloneqq \langle |\nabla_s \log p(s|y)|^2 \rangle_{p(s,y)}$, and the corresponding information potential by $\Phi[s|y] \coloneqq H[s|y] + \frac{N_s}{2} \log \frac{J[s|y]}{2\pi e N_s}$, where $N_s = \dim(s)$. The conditional densities are assumed to be twice continuously differentiable and sufficiently regular to justify the differentiation and integration operations used below.

In addition to the entropic contribution fixed in **SI.7**, the same Markov structure constrains the Fisher term. For the Markov chain $s \to o \to x$, $p(s|x) = \int p(s|o)p(o|x)do$ holds. Differentiating with respect to $s$ gives $\nabla_s \log p(s|x) = \langle \nabla_s \log p(s|o) \rangle_{p(o|s,x)}$, where $p(o|s,x) = p(s|o)p(o|x)/p(s|x)$ was used. Conditional Jensen's inequality therefore implies $|\nabla_s \log p(s|x)|^2 \le \langle |\nabla_s \log p(s|o)|^2 \rangle_{p(o|s,x)}$. Taking expectations over $p(s,x)$ yields

$$J[s|x] \le J[s|o]. \tag{42}$$

Thus, posterior sharpness about $s$ cannot increase downstream along the Markov chain. Combining this Fisher bound with the entropy inequality $H[s|x] \le H[s|o] + H[o|x] - H_o^*$ provides

$$0 \le \Phi[s|x] \le \Phi[s|o] + H[o|x] - H_o^* \le \Phi[s|o] + H[o|o'] - H_o^*. \tag{43}$$

The lower bound follows from Stam's inequality, with equality only for isotropic Gaussian conditional structure.

A complementary bound is obtained for the conditional structure of $x$. Define $J[x|y] \coloneqq \langle |\nabla_x \log p(x|y)|^2 \rangle_{p(x,y)}$, and $\Phi[x|y] \coloneqq H[x|y] + \frac{N_x}{2} \log \frac{J[x|y]}{2\pi e N_x}$, where $N_x = \dim(x)$. From $s \to o \to x$, $p(x|s) = \int p(x|o)p(o|s)do$ holds. Differentiating with respect to $x$ gives $\nabla_x \log p(x|s) = \langle \nabla_x \log p(x|o) \rangle_{p(o|s,x)}$, where $p(o|s,x) = p(x|o)p(o|s)/p(x|s)$ was used. Conditional Jensen's inequality then gives

$$J[x|s] \leq J[x|o]. \tag{44}$$

Together with the entropy inequality $H[x|s] \leq H[x|o] + H[o|s] - H_o^*$, this yields

$$0 \leq \Phi[x|s] \leq \Phi[x|o] + H[o|s] - H_o^*. \tag{45}$$

Equations (43) and (45) show that information potential is constrained by the same Markov structure that bounds conditional entropy. The first inequality states that the structural complexity of $p(s|x)$ is bounded by the intrinsic difficulty of external-state inference, $\Phi[s|o]$, together with representational or reconstructive loss. Hence, if $\Phi[s|o]$ and $H[o|o']$ are small, the posterior induced by the internal state remains close to a minimal-complexity form. The second inequality gives the complementary constraint. The structural complexity of $p(x|s)$ is bounded by the intrinsic complexity of the encoder, $\Phi[x|o]$, and by the ambiguity of observation generation, $H[o|s]$. Thus, if $p(x|o)$ is geometrically simple and observations are only weakly ambiguous given external states, $p(x|s)$ cannot become arbitrarily complex. Therefore, when these conditions hold, internal representations can preserve both the information about external causes and the geometric simplicity of the corresponding posterior.

Together, these inequalities furnish a dual criterion for representational quality. A good internal code should preserve externally relevant information while maintaining simple conditional structure in both directions of the $s \leftrightarrow x$ relation. Among representations that preserve comparable information, $\Phi$ distinguishes those that induce smooth and tractable posterior landscapes from those that retain unnecessary ruggedness, anisotropy, or multimodality.

## SI.9 Exact rigidity of reciprocal Gaussian conditionals

This section derives an exact rigidity result. When both $\Phi[s|x]$ and $\Phi[x|s]$ vanish, both conditional distributions are isotropic Gaussians. This constrains the joint surprisal, $-\log p(s,x)$, to be quadratic and the map between $s$ and $x$ to be affine.

Let $s \to o \to x \to o'$ be the Markov chain defined above. Consider the exact case $\Phi[x|s] = \Phi[s|x] = 0$. Both conditional distributions then attain equality in the information-potential bound. Hence $x|s$ and $s|x$ are isotropic Gaussians with constant scales. The forward conditional can therefore be written as $x = f(s) + \xi$ with $\xi \sim \mathcal{N}[0, \sigma_x^2 I]$, so that the joint surprisal is $\frac{1}{2\sigma_x^2}|x - f(s)|^2$, up to an additive term depending only on $s$. Because $s|x$ is also isotropic Gaussian, the same joint surprisal must be quadratic in $s$ for each fixed $x$. Under the regularity and non-degeneracy assumptions used above, this is possible only when $f(s)$ is affine.

To see this, let $p(s,x)$ be a positive twice-differentiable density, and suppose that $p(x|s) = \mathcal{N}[f(s), \sigma_x^2 I]$ and $p(s|x) = \mathcal{N}[g(x), \sigma_s^2 I]$, with fixed $\sigma_x^2$ and $\sigma_s^2$. Then, one obtains

$$-\log p(s,x) = \frac{1}{2\sigma_x^2}|x - f(s)|^2 + \alpha(s) = \frac{1}{2\sigma_s^2}|s - g(x)|^2 + \beta(x). \tag{46}$$

where $\alpha(s)$ and $\beta(x)$ collect the additive terms depending only on $s$ and $x$, respectively. Taking the mixed derivatives provide $-\nabla_s \nabla_x^{\mathrm{T}} \log p(s,x) = \nabla_s \{\sigma_x^{-2}(x - f)\}^{\mathrm{T}} = -\sigma_x^{-2}(\nabla_s f)^{\mathrm{T}}$ and $-\nabla_x \nabla_s^{\mathrm{T}} \log p(s,x) = \nabla_x \{\sigma_s^{-2}(s - g)\}^{\mathrm{T}} = -\sigma_s^{-2}(\nabla_x g)^{\mathrm{T}}$, where $\nabla_s$ and $\nabla_x$ denote the *s*- and *x*-derivatives, respectively. Thus, the cross terms that depend jointly on $s$ and $x$ must satisfy

$$\sigma_x^{-2} \nabla_s f(s) = \sigma_s^{-2} \left(\nabla_x g(x)\right)^{\mathrm{T}}, \tag{47}$$

for every $(s, x)$. The left-hand side depends only on $s$, whereas the right-hand side depends only on $x$. Since they are equal for every $(s, x)$, $\nabla_s f(s)$ and $\nabla_x g(x)$ must be constant. Hence, $f(s)$ and $g(x)$ are affine.

Writing $f(s) = a + As$ and $g(x) = b + Bx$ with matrices $A$ and $B$ and translation vectors $a$ and $b$, the two expressions for $-\log p(s, x)$ have the same bilinear term, $-x^{\mathrm{T}} As/\sigma_x^2$. After cancelling this term, the remaining equality separates into a function of $x$ and a function of $s$. Equality for all $(s, x)$ then forces $\alpha(s)$ and $\beta(x)$ to be quadratic. Thus, $-\log p(s, x)$ is quadratic in $(s, x)$. Normalisability therefore implies that $p(s, x)$ is jointly Gaussian.

Thus, in the exact zero-potential limit, reciprocal Gaussian minimality restricts the conditional means to affine maps and the joint surprisal to a quadratic form. If, in addition, the reverse conditional precision is dominated by the likelihood curvature at the scale considered, bidirectional isotropy imposes $A^{\mathrm{T}} A = (\sigma_x^2/\sigma_s^2) I$. For $N_s = N_x$, this provides $A = (\sigma_x/\sigma_s) R$, where $R$ is an orthogonal matrix. Hence, in the likelihood-dominated isotropic limit, the admissible $s \leftrightarrow x$ relation collapses to a scaled rotation and translation, $f(s) = (\sigma_x/\sigma_s) Rs + a$. Linearity therefore follows from reciprocal Gaussian-minimal structure, whereas conformality follows from the additional bidirectional isotropy of the conditional geometry.

## SI.10 Approximate rigidity of reciprocal Gaussian conditionals

This section derives an approximate rigidity result. When both $\Phi[s|x]$ and $\Phi[x|s]$ are small, both conditional geometries are close to their minimal forms. This constrains the joint surprisal $-\log p(s, x)$ to be close to quadratic and the map between $s$ and $x$ to be close to affine.

The approximate case follows the same structure as the exact case in **SI.9**, but replaces exact Gaussian minimality by a local small-noise approximation. Suppose that $H[x|s]$ is small and the conditional mean is smooth, so that $x$ is concentrated around a differentiable map of $s$. A general stochastic transformation $x = f(s, \xi_0)$ can then be expanded to first order in a small Gaussian noise variable $\xi_0$, $x = f(s, 0) + (\nabla_{\xi_0} f)(s, 0)\xi_0 + \mathcal{O}(|\xi_0|^2)$. Absorbing the resulting local noise gain into the state-dependent covariance $\Sigma_x(s)$ yields a general small-noise forward map, to leading order in the noise variable, $x = f(s) + \xi$ with $\xi \sim \mathcal{N}[0, \Sigma_x(s)]$. For this conditional family,

$$\Phi[x|s] = \frac{1}{2}\langle \log|\Sigma_x(s)|\rangle_{p(s)} + \frac{N_x}{2}\log\left(\frac{1}{N_x}\langle \operatorname{tr}\Sigma_x(s)^{-1}\rangle_{p(s)}\right), \tag{48}$$

where $N_x = \dim(x)$. Defining $\gamma(s) \coloneqq \frac{1}{N_x}\operatorname{tr}\Sigma_x(s)^{-1}$, this can be decomposed as

$$\Phi[x|s] = \frac{N_x}{2}\left\langle \log\gamma(s) - \frac{1}{N_x}\log|\Sigma_x(s)^{-1}|\right\rangle_{p(s)} + \frac{N_x}{2}\left(\log\langle\gamma(s)\rangle_{p(s)} - \langle\log\gamma(s)\rangle_{p(s)}\right). \tag{49}$$

Both terms are non-negative by Jensen's inequality. The first term measures local anisotropy of the conditional noise by applying Jensen's inequality—or the arithmetic–geometric mean inequality—to the eigenvalues of $\Sigma_x(s)^{-1}$ at fixed $s$, and vanishes only when these eigenvalues are all equal, i.e., when $\Sigma_x(s)$ is isotropic. The second term applies Jensen's inequality to the scalar precision $\gamma(s)$ to measure variation of scale across $s$, and vanishes only when $\gamma(s)$ is independent of $s$.

Therefore, $\Phi[x|s] = \mathcal{O}(\delta^2)$ implies that, for $p(s)$-typical states in the connected region where the small-noise expansion is valid, the conditional covariance is isotropic with an $s$-independent scale up to $\mathcal{O}(\delta)$,

$$\Sigma_x(s) = \sigma_x^2 I + \mathcal{O}(\delta), \tag{50}$$

where $\sigma_x^2$ is a non-zero, $s$-independent variance. Here, $p(s)$-typical means that, after excluding a set with any prescribed small probability mass, the covariance deviation is $\mathcal{O}(\delta)$ over the remaining states, with the proportionality constant allowed to depend on the chosen probability threshold. The $\mathcal{O}(\delta)$ scaling follows from the quadratic expansion of $\Phi[x|s]$ around its isotropic fixed-scale minimum.

The reverse conditional imposes an independent geometric constraint. Suppose that $f(s)$ is locally invertible, the curvature of the prior over $s$ is lower order than the likelihood curvature, and the noise $\xi$ is small. Under these assumptions, the likelihood dominates locally, so the leading-order reverse fluctuations are obtained by propagating the small forward noise through the inverse map. For fixed $x$, $s = f^{-1}(x-\xi) = f^{-1}(x) - \nabla_x f^{-1}(x)\xi + \mathcal{O}(|\xi|^2) = \hat{s}(x) + \eta$ holds, where $\hat{s}(x) = f^{-1}(x)$ and $\eta = -\nabla_x f^{-1}(x)\xi + \mathcal{O}(|\xi|^2)$. Hence, the local covariance of $s|x$ satisfies $\mathrm{Cov}[s|x] = \Sigma_s(x) = \nabla_x f^{-1}(x)\Sigma_x(\hat{s})\nabla_x f^{-1}(x)^{\mathrm{T}} + \mathcal{O}(\delta)$, or equivalently, $\Sigma_s(x)^{-1} = \nabla_s f(\hat{s})^{\mathrm{T}}\Sigma_x(\hat{s})^{-1}\nabla_s f(\hat{s}) + \mathcal{O}(\delta)$. Same as above, if $\Phi[s|x] = \mathcal{O}(\delta^2)$, the reverse conditional covariance must also be almost isotropic with almost constant scale,

$$\Sigma_s(x) = \sigma_s^2 I + \mathcal{O}(\delta), \tag{51}$$

with a non-zero, $x$-independent variance $\sigma_s^2$. The same typical-state qualification applies here with respect to $p(x)$. It follows $\nabla_s f(s)^{\mathrm{T}}\Sigma_x(s)^{-1}\nabla_s f(s) = \sigma_s^{-2} I + \mathcal{O}(\delta)$. Combining this with equation (50) yields $\nabla_s f(s)^{\mathrm{T}}\nabla_s f(s) = (\sigma_x^2/\sigma_s^2)I + \mathcal{O}(\delta)$. Thus, up to a uniform scale, the Jacobian of $f(s)$ is close to an orthogonal matrix.

In essence, small $\Phi[x|s]$ enforces an almost isotropic, fixed-scale forward noise, whereas small $\Phi[s|x]$ enforces an almost isotropic, fixed-scale reverse posterior geometry. Together they imply that, on the same typical-state region, the Jacobian of $f(s)$ is close to a scaled orthogonal matrix.

It remains to show that the Jacobian of $f(s)$ cannot vary freely with $s$. Its derivation follows from the same reciprocal-consistency argument as in the exact case in **SI.9**. Within the local Gaussian approximation above, the joint surprisal admits two equivalent representations,

$$-\log p(s,x) = \frac{1}{2}\big(x - f(s)\big)^{\mathrm{T}}\Sigma_x(s)^{-1}\big(x - f(s)\big) + \alpha(s)$$

$$= \frac{1}{2}\big(s - g(x)\big)^{\mathrm{T}}\Sigma_s(x)^{-1}\big(s - g(x)\big) + \beta(x), \tag{52}$$

where $\alpha(s)$ and $\beta(x)$ denote collected additive terms. This is a straightforward generalisation of equation (46) to state-dependent Gaussian neighbourhoods. Taking the mixed derivatives provides $-\nabla_s\nabla_x^{\mathrm{T}}\log p(s,x) = \nabla_s\big\{\Sigma_x(s)^{-1}\big(x - f(s)\big)\big\}^{\mathrm{T}}$ and $-\nabla_x\nabla_s^{\mathrm{T}}\log p(s,x) = \nabla_x\big\{\Sigma_s(x)^{-1}\big(s - g(x)\big)\big\}^{\mathrm{T}}$. Thus, the cross terms that depend jointly on $s$ and $x$ must satisfy

$$\nabla_s\big\{\Sigma_x(s)^{-1}\big(x - f(s)\big)\big\} = \nabla_x\big\{\Sigma_s(x)^{-1}\big(s - g(x)\big)\big\}^{\mathrm{T}}, \tag{53}$$

for every $(s,x)$. Taking one further derivative with respect to both $s$ and $x$ eliminates the terms involving $f$ and $g$ and gives, componentwise, $\partial_{s_l}\partial_{s_k}[\Sigma_x(s)^{-1}]_{ij} = \partial_{x_j}\partial_{x_i}[\Sigma_s(x)^{-1}]_{kl}$. Since the two sides depend only on $s$ and $x$, respectively, both must be constant. Hence, the precision matrices $\Sigma_x(s)^{-1}$ and $\Sigma_s(x)^{-1}$ are at most quadratic functions of $s$ and $x$, respectively.

Equations (50) and (51), together with finite non-zero conditional variances, imply $\Sigma_x(s)^{-1} - \sigma_x^{-2}I = \mathcal{O}(\delta)$ and $\Sigma_s(x)^{-1} - \sigma_s^{-2}I = \mathcal{O}(\delta)$ on $\mathcal{O}(1)$ connected regions of $p(s)$- and $p(x)$-typical states, respectively. Since these deviations are at most quadratic and remain $\mathcal{O}(\delta)$ over $\mathcal{O}(1)$ ranges of $s$ and $x$, their linear and quadratic coefficients must also be $\mathcal{O}(\delta)$. It follows that $\nabla_s\Sigma_x(s)^{-1} = \mathcal{O}(\delta)$ and $\nabla_x\Sigma_s(x)^{-1} = \mathcal{O}(\delta)$. Returning to equation (53), the terms containing these precision gradients are therefore $\mathcal{O}(\delta)$, yielding

$$\nabla_s\{\Sigma_x(s)^{-1}f(s)\} - \nabla_x\{\Sigma_s(x)^{-1}g(x)\}^{\mathrm{T}} = \mathcal{O}(\delta). \tag{54}$$

The remaining terms depend only on $s$ and $x$, respectively. Since they agree up to $\mathcal{O}(\delta)$ for every $(s,x)$, there exist constant matrices $A$ and $B$ such that $\nabla_s\{\Sigma_x(s)^{-1}f(s)\} = \sigma_x^{-2}A + \mathcal{O}(\delta)$ and $\nabla_x\{\Sigma_s(x)^{-1}g(x)\} = \sigma_s^{-2}B + \mathcal{O}(\delta)$.

Integration over any connected region on which these residual bounds hold then provides $\Sigma_x(s)^{-1}f(s) = \sigma_x^{-2}(a + As) + \mathcal{O}(\delta)$ and $\Sigma_s(x)^{-1}g(x) = \sigma_s^{-2}(b + Bx) + \mathcal{O}(\delta)$. From equations (50) and (51), it follows $f(s) = a + As + \mathcal{O}(\delta)$ and $g(x) = b + Bx + \mathcal{O}(\delta)$. Substituting these relations and equations (50) and (51) back into equation (52) shows that $\alpha(s)$ and $\beta(x)$ are quadratic up to $\mathcal{O}(\delta)$, and hence that the joint surprisal is quadratic up to $\mathcal{O}(\delta)$ on the same region. Combining this reciprocal-consistency constraint with $\nabla_s f(s)^{\mathrm{T}}\nabla_s f(s) = (\sigma_x^2/\sigma_s^2)I + \mathcal{O}(\delta)$ provides $A^{\mathrm{T}}A = (\sigma_x^2/\sigma_s^2)I + \mathcal{O}(\delta)$. For $N_s = N_x$, this implies $A = (\sigma_x/\sigma_s)R + \mathcal{O}(\delta)$, where $R$ is an orthogonal matrix.

In summary, approximate bidirectional minimality of $\Phi$ imposes geometric rigidity on the conditional relation between $s$ and $x$. $\Phi[x|s]$ controls anisotropy and scale variation in the forward conditional noise, whereas $\Phi[s|x]$ controls anisotropy and scale variation in the reverse posterior geometry. When both are small, the two conditional descriptions become approximately isotropic with fixed scale, and their reciprocal consistency prevents the local scaled rotation from varying freely with $s$. Consequently, the admissible relation is constrained to an approximately conformal affine map, with deviations of order $\mathcal{O}(\delta)$ when $\Phi[s|x] + \Phi[x|s]$ is of order $\mathcal{O}(\delta^2)$. In the exact limit $\delta \to 0$, the residuals vanish and reciprocal Gaussian minimality yields affine conditional means and a quadratic joint surprisal. Under the small-noise likelihood-dominated condition, this affine relation collapses to a fixed scaled rotation and translation, $f(s) = (\sigma_x/\sigma_s)Rs + a$. Representation inequalities therefore define the law-level envelope within which geometric rigidity arises.

**References**

1. Bayes, T. LII. An essay towards solving a problem in the doctrine of chances. By the late Rev. Mr. Bayes, FRS communicated by Mr. Price, in a letter to John Canton, AMFR S. *Philos. Trans. R. Soc. Lond.* **53**, 370–418 (1763).

2. Stam, A. J. Some inequalities satisfied by the quantities of information of Fisher and Shannon. *Inf. Comput.* **2**, 101–112 (1959).

3. Bochner, S. Vector fields and Ricci curvature. *Bull. Am. Math. Soc.* **52**, 776–797 (1946).

4. Cover, T. M. & Thomas, J. A. *Elements of informaRon theory.* John Wiley & Sons (1991).